\documentclass[12pt]{article}
\usepackage{graphicx}
\usepackage{cite}
\usepackage{color}
\newcommand{\One}{1\hspace{-1.6mm}1}

\newcommand{\tso}{$^{3\!}S_1$}
\newcommand{\ontso}{$1\,{}^{3\!}S_1$}
\newcommand{\twtso}{$2\,{}^{3\!}S_1$}
\newcommand{\thtso}{$3\,{}^{3\!}S_1$}
\newcommand{\fotso}{$4\,{}^{3\!}S_1$}
\newcommand{\tdo}{$^{3\!}D_1$}
\newcommand{\ontdo}{$1\,{}^{3\!}D_1$}
\newcommand{\twtdo}{$2\,{}^{3\!}D_1$}
\newcommand{\thtdo}{$3\,{}^{3\!}D_1$}
\newcommand{\tpo}{$^{3\!}P_1$}
\newcommand{\spo}{$^{1\!}P_1$}
\begin{document}
\title{Unquenched quark-model calculation of excited $\rho$
resonances and $P$-wave $\pi\pi$ phase shifts\footnote{Talk given by G.~Rupp
at the Mini-Workshop Bled 2015, \em ``Exploring Hadron Resonances'', \em Bled,
Slovenia, 5--11 July 2015.}} 
\author{
Susana Coito \\
{\normalsize\it Institute of Modern Physics, CAS, Lanzhou 730000, China} \\
{\small susana@impcas.ac.cn} \\[.3cm]
George Rupp \\
{\normalsize\it CeFEMA, Instituto Superior T\'{e}cnico} \\
{\normalsize\it Universidade de Lisboa, 1049-001 Lisboa, Portugal}\\
{\small george@ist.utl.pt} \\[.3cm]
Eef van Beveren\\
{\normalsize\it Centro de F\'{\i}sica Computacional, Departamento de
F\'{\i}sica} \\
{\normalsize\it Universidade de Coimbra, 3004-516 Coimbra, Portugal}\\
{\small eef@teor.fis.uc.pt}
\\[1cm]
}

\maketitle

\begin{abstract}
The $\rho(770)$ vector resonance, its radial recurrences, and the corresponding
$P$-wave $\pi\pi$ phase shifts are investigated in an unquenched quark model
with all classes of relevant decay channels included, viz.\
pseudoscalar-pseudoscalar, vector-pseudoscalar, vector-vector,
vector-scalar, axialvector-pseudoscalar, and axialvector-vector,
totalling 26 channels. Two of the few model parameters are fixed at previously
used values, whereas the other three are adjusted to the $\rho(770)$ resonance
and the lower $P$-wave $\pi\pi$ phases. Preliminary results indicate the
model's capacity to reproduce these phases as well as the $\rho$ mass and
width. However, at higher energies the phase shifts tend to rise too sharply.
A possible remedy is an extension of the model so as to handle resonances in
the final states for most of the included decay channels. Work in progress.
\end{abstract}

\section{Introduction}
The radial recurrences of the $\rho(770)$ vector resonance play a crucial role
in light-meson spectroscopy, owing to the several observed states in the
PDG tables \cite{PDG2014}, up to 1.9 GeV, and the corresponding $P$-wave
pion-pion phase shifts and inelasticities measured in several experiments
\cite{PDG2014}. These resonances may shed a lot of light on the underlying
quark-confinement force as well as the strong-decay mecanism, both assumed to
result from low-energy QCD. However, there are two serious problems, one
experimental and the other theoretical. First of
all, excited $\rho$ states listed in the PDG tables are far from well
established, even the generally undisputed $\rho(1450)$ \cite{PDG2014}
resonance. For example, under the entry of the latter state in the PDG meson
listings, experimental observations have been collected with masses in the
range 1250--1582~MeV. The state of lowest mass here corresponds to the
$\omega\pi^0$ resonance at 1.25 GeV first observed by Aston {\it et al.} \/in
1980 \cite{PLB92p211}. Several other experiments have confirmed such a vector
$\rho^\prime$ resonance in the mass interval 1.25--1.3~GeV, both in the
$\omega\pi^0$ \cite{rho1250omegapi} and $\pi\pi$ \cite{rho1250pipi} channels.
More recently, a combined $\mathcal{S}$-matrix and Breit-Wigner (BW) analysis
\cite{NPA807p145}, using $P$-wave $\pi\pi$ data from the early 1970s, not only
confirmed a $\rho(1250)$, but even found it to be much more important for a
good fit to the data than the $\rho(1450)$.
In view of these findings, it is simply inconceivable that no separate
$\rho(1250)$ entry has been created in the PDG tables, and to make things
worse, the latter analysis \cite{NPA807p145} is not even included in the PDG
references. The reason for these omissions seems to be based on theory bias,
which is the second problem.
Indeed, the renowned relativised quark model for mesons by Godfrey \& Isgur
\cite{PRD32p189} predicted the first radial $\rho$ excitation at 1.45~GeV. As a
matter of fact, practically all quark models employing the usual
Coulomb-plus-linear confining potential find a $\rho^\prime$ at about the same
mass and cannot accommodate a $\rho(1250)$ (see, however,
Ref.~\cite{PRD27p1527}). 

Unfortunately, no
fully unquenched lattice calculations including both $q\bar{q}$ and two-meson
interpolating fields have been carried out so far beyond the ground-state
$\rho(770)$ \cite{PRD84p054503}. Nevertheless, in the strange-meson sector
such a calculation was done recently \cite{PRD88p054508}, reproducing the
$K^\star(892)$ resonance in $P$-wave elastic $K\pi$ scattering, with mass and
width close to the experimental values. Moreover, the first radial excitation
was identified as well, tentatively at 1.33~GeV, which should correspond to the
$K^\star(1410)$ \cite{PDG2014} resonance. The latter is thus determined
on the lattice to be a normal quark-antiquark resonance. From simple quark-mass
considerations, one is led to conclude the same for the $\rho^\prime$
originally found by Aston {\it et al.} \/\cite{PLB92p211} at 1.25~GeV, which
contradicts the speculation in Ref.~\cite{ZPC60p187} that it \em ``has
necessarily to be an exotic''. \em Note that including meson-meson
interpolators is absolutely crucial to reliably predict the mass of an excited 
resonance like the ${K^{\star}}^\prime$, since an also unquenched lattice
calculation without considering decay, by partly the same authors
\cite{POSHADRON2013p118}, predicted a ${K^{\star}}^\prime$ mass a full 300 MeV
heavier than in Ref.~\cite{PRD88p054508}.

In the present work, we analyse the issue of radial $\rho$ recurrences by
attempting to describe the elastic and inelastic $P$-wave $\pi\pi$ phase shifts
in the context of an unquenched quark model that has been successfully applied
to several problematic mesons (see e.g.\ Ref.~\cite{APPBPS8p139} for a very
brief recent review). In this so-called Resonance-Spectrum-Expansion (RSE) 
model, the manifest non-perturbative inclusion of all relevant two-meson
channels alongside a confined $q\bar{q}$ sector allows for a phenomenlogical
decription of excited meson resonances in the same spirit as the referred
lattice calculation \cite{PRD88p054508}. An additional advantage is that the
RSE model yields an exactly unitary and analytic $\mathcal{S}$-matrix for
any number of included quark-antiquark and two-meson channels, whereas the
lattice still faces serious problems in the case of inelastic resonances and
highly excited states.

\section{RSE modelling of \boldmath{$P$}-wave \boldmath{$\pi\pi$} scattering}

The general expressions for the RSE off-energy-shell $\mathcal{T}$-matrix and
corresponding on-shell $\mathcal{S}$-matrix have been given in several
papers (see e.g.\ Ref.~\cite{PRD84p094020}). In the present case of $P$-wave
$\pi\pi$ scattering, the quantum numbers of the system are
$I^GJ^{PC}=1^+\,1^{--}$, which couples to the $I\!=\!1$ quark-antiquark state
$(u\bar{u}-d\bar{d})/\sqrt{2}$ in the spectroscopic channels \tso\ and \tdo.
In the meson-meson sector, we only consider channels allowed by total angular
momentum $J$, isospin $I$, parity $P$, and G-parity $G$. The included
combinations are pseudoscalar--pseudoscalar (PP),
vector-pseudoscalar (VP), vector-vector (VV), vector-scalar (VS),
axialvector-pseudoscalar (AP), and axialvector-vector (AV),
with mesons from the lowest-lying pseudoscalar, vector, scalar, and axialvector
nonets listed in the PDG \cite{PDG2014} tables. Here,  ``axialvector'' may 
refer to $J^{PC}=1^{++}$, $J^{PC}=1^{+-}$, or $J^P=1^+$ for mesons with no
definite $C$-parity. This choice of meson-meson channels is motivated by the
observed two- and multi-particle decays of the $\rho$ recurrences up to the
$\rho(1900)$ \cite{PDG2014}, which include several intermediate states 
containing resonances from the referred nonets. For instance, the PDG lists
\cite{PDG2014} under the $4\pi$ decays of the $\rho(1450)$ the modes
$\omega\pi$, $a_1(1260)\pi$, $h_1(1170)\pi$, $\pi(1300)\pi$, $\rho\rho$, and
$\rho(\pi\pi)_{\mbox{\scriptsize $S$-wave}}$, where
$(\pi\pi)_{\mbox{\scriptsize $S$-wave}}$ is probably dominated by the
$f_0(500)$ \cite{PDG2014} scalar resonance. By the same token, the $6\pi$
decays of the $\rho(1900)$ will most likely include important contributions
from modes as $b_1(1235)\rho$, $a_1(1260)\omega$, \ldots\ . For consistency of
our calculation, we generally include complete nonets in the allowed decays,
and not just individual modes observed in experiment. The only exception is the
important $\pi(1300)\pi$ P$^\prime$P mode, because no complete nonet of
radially excited pseudoscalar mesons has been observed so far \cite{PDG2014}.
The included 26 meson-meson channels are given in Table~1. \\

\begin{table}[h]
\begin{center}
\begin{tabular}{|c|l|c|}
\hline
Nonets & \hspace*{45mm}Two-Meson Channels &  $L$ \\ \hline\hline
PP & $\pi\pi$, $KK$ & 1 \\ \hline
VP & $\omega\pi$, $\rho\eta$, $\rho\eta^\prime$, $K^\star K$ & 1 \\ \hline
VV & $\rho\rho$, $K^\star K^\star$  & 1 \\ \hline
VS & $\rho f_0(500)$, $\omega a_0(980)$, $K^\star K_0^\star(800)$ & $0,2$ \\ 
\hline
AP & $a_1(1260)\pi$, $b_1(1235)\eta$, $b_1(1235)\eta^\prime$,
     $h_1(1170)\pi$, $K_1(1270)K$, $K_1(1400)K$ & 0 \\ \hline
AV & $a_1(1260)\omega$, $b_1(1235)\rho$, $f_1(1285)\rho$,
     $K_1(1270)K^\star$, $K_1(1400)K^\star$ & 0 \\ \hline
P$^\prime$P & $\pi(1300)\pi$ & 1 \\ \hline
\end{tabular}
\end{center}
\caption{Included classes of decay channels containing mesons listed
\cite{PDG2014} in the PDG tables, with the respective orbital angular momenta.
Note that the included P$^\prime$P decay mode is incomplete (see text
above).}
\label{channels}
\end{table}
Notice that the VS channels count twice, because they can have $L\!=\!0$ or
$L\!=\!2$. However, the $S$-wave only couples to the \tso\ $q\bar{q}$
channel and the $D$-wave only to the \tdo. The relative couplings between the
$q\bar{q}$ and meson-meson channels are determined using the scheme of
Ref.~\cite{ZPC21p291}, based on overlaps of  harmonic-oscillator (HO) wave
functions. Special care is due in the cases of flavour mixing
($\eta,\eta^\prime$), and mixing of the $C$-parity eigenstates \tpo\ and \spo\
\cite{PRD84p094020}, as the strange axialvector mesons $K_1(1270)$ and
$K_1(1400)$ have no definite $C$-parity.

Coming now to describing the data, we have to adjust the model parameters.
Two of these, namely the non-strange constituent quark mass and the HO
oscillator frequency, are as always fixed \cite{PRD84p094020} at the values
$m_n=m_u=m_d=406$~MeV and $\omega=190$~MeV. This yields a largely degenerate
bare $\rho$ spectrum with energy levels 1097~MeV (\ontso), 1477~MeV
(\twtso/\ontdo), 1857~MeV (\thtso/\twtdo), 2237~MeV (\fotso/\thtdo),
\ldots\ . This bare spectrum is then deformed upon unquenching, that is, by
allowing $q\bar{q}$ pair creation. This results in real or complex mass
shifts due to meson-loop contributions, depending on decay channels being closed
or open, respectively. Note that these shifts are non-perturbative and can only
be determined, for realistic coupling strengths, by numerically finding the
poles of the $\mathcal{S}$-matrix. The adjustable parameters are the overall
dimensionless coupling constant $\lambda$, the ``string-breaking'' radius
$r_0$ for transitions between the $q\bar{q}$ and two-meson channels, and a
range parameter $\alpha$ for weakening subthreshold contributions via a form
factor. The coupling $\lambda$ is usually in the range 3--5, $r_0$ should be
of the order of 1~fm for systems made of light quarks (with mass $m_n$), and
the $\alpha$ value used in several previous papers is 4~GeV$^{-2}$ (see e.g.\
Ref.~\cite{PRD80p094011}).

In view of these strong limitations, it is quite remarkable that we can 
obtain a good fit to the $P$-wave $\pi\pi$ phase shifts up to 1.2~GeV with the
choice $\lambda=5.3$, $r_0=0.9$~fm, and $\alpha=4$~GeV$^{-2}$. Moreover, the
corresponding $\rho(770)$ pole comes out at the very reasonable energy
$E=(754-i67)$~MeV. At higher energies, though, the $\pi\pi$ phases tend to 
rise too fast and no good description has been obtained so far, also due to the
very little fitting freedom. As for the poles of the higher $\rho$ recurrences,
we find at least four in the energy range 1.2--2.0 GeV, in 
agreement with the PDG \cite{PDG2014} and also Ref.~\cite{NPA807p145}, albeit
at yet quite different energies. In particular, there are two poles between 1.2
and 1.5~GeV, in qualitative agreement with Ref.~\cite{NPA807p145}. However,
these pole positions are extremely sensitive to the precise values of the
parameteres $\lambda$, $r_0$, and $\alpha$, so that they should not be taken at
face value as long as no good fit is achieved of the observables above 1.2~GeV.

A possibility to improve the fit is by allowing 
different decay radii for the several classes of decay channels (PP, VP, VV,
VS, AP, AV, P$^\prime$P), which would be logical in view of the detailed,
channel-dependent transition potentials derived in Ref.~\cite{ZPC21p291}. Such
an additional flexibility will neither affect the exact solvability of the RSE
$\mathcal{T}$-matrix, nor its analyticity.

An addditional possible model extension we shall discuss in the next section.
\section{Resonances in asymptotic states}
The decay channels listed in Table~1 contain several resonances, some of
which are even extremely broad, with widths exceeding 300~MeV. So treating
the corresponding thresholds as being sharp, at well-defined real energies, is
certainly an approximation, which may produce too sudden effects at
threshold openings. Ideally, one would like to describe a resonance
in the final state via a smooth function of real energy, corresponding to an
experimental cross section in which the resonance is observed. By discretising
such a function, one could in principle  describe each final-state resonance 
through a large number of effective thresholds. However, this would lead to
a proliferation of channels and to a true explosion of Riemann sheets,
making the tracing of complex poles impracticable. 

An alternative way to handle a resonance in asymptotic states is to replace
its real mass by a complex one, on the basis of the PDG \cite{PDG2014}
resonance mass and total width. However, this inevitably destroys unitarity
of the $\mathcal{S}$-matrix. Nevertheless, its symmetry will be unaffected, 
which can be used to define a new matrix that is unitary again, and so take
over the role of $\mathcal{S}$. Here, we closely follow the derivation given
in Ref.~\cite{EPJC71p1762}.

An arbitrary symmetric matrix $\mathcal{S}$ can be decomposed, via Takagi
\cite{JJM1p82} factorisation, as
\begin{equation}
\mathcal{S} \; = \; VDV^{T} \; ,
\label{takagi}
\end{equation}
where $V$ is unitary and $D$ is a real non-negative diagonal matrix.
Then we get
\begin{equation}
\mathcal{S}^\dag\mathcal{S} = (V^T)^\dag DV^\dag VDV^T = (V^T)^\dag D^2V^T
= U^\dag D^2U ,
\label{sdags}
\end{equation}
where we have defined $U\equiv V^T$, which is obviously unitary, too. So the
diagonal elements of $D=\sqrt{U\mathcal{S}^\dag\mathcal{S}U^\dag}$  are the
square roots of the eigenvalues of the positive Hermitian matrix
$\mathcal{S}^\dag\mathcal{S}$, which are all real and non-negative. Moreover,
since $\mathcal{S}=\One+2i\mathcal{T}$ is manifestly non-singular, the
eigenvalues of $\mathcal{S}^\dag\mathcal{S}$ are even all
non-zero and $U$ is unique. Thus, we may define
\begin{equation}
\mathcal{S}^\prime \; \equiv \; \mathcal{S}U^\dag D^{-1}U \; .
\label{sprime}
\end{equation}
Then, using Eq.~(\ref{takagi}) and $V=U^T$, we have
\begin{equation}
\mathcal{S}^\prime \; = \; U^TDUU^\dag D^{-1}U \; = \; U^TU \; ,
\label{sprimesym}
\end{equation}
which is obviously symmetric. But it is also unitary, as
\begin{equation}
(U^TU)^\dag=U^\dag(U^\dag)^T=U^{-1}(U^{-1})^T=(U^TU)^{-1}\;.
\label{sprimeunit}
\end{equation}
So $\mathcal{S}^\prime$ has the required properties to be defined as the
scattering matrix for a process with complex masses in the asymptotic states.
Note that this empirical method has been applied very successfully to the
enigmatic $X(3872)$ charmonium state in Ref.~\cite{EPJC71p1762}.

\section{Conclusions}
We have presented preliminary results of an unquenched quark-model study aimed
at determining the complex pole positions of vector $\rho$ recurrences up to
2~GeV, motivated by the poor status of these resonances in the PDG tables
\cite{PDG2014} and their importance for light-meson spectroscopy. The employed
RSE model was applied in the past to a variety of problematic mesons,
with very good results \cite{APPBPS8p139}. The here included classes of
meson-meson channels cover most of the observed strong decays. The three
adjustable model parameters were fitted to the $P$-wave $\pi\pi$ phase shifts
up to about 1.2~GeV, allowing a good reproduction of these data and a very 
reasonable $\rho(770)$ resonance pole position.

However, at energies above 1.2~GeV the thus calculated phases rise too fast,
and a globally good fit including the higher $\pi\pi$ phases is not feasible
with only three parameters. A possible model extension amounts to allowing
different decay radii for the different classes of meson-meson channels, which
will not spoil the nice model features. Another extension to be considered is
the use of complex physical masses for the final-state resonances in nearly all
channels of Table~1. This will require a redefinition of the
$\mathcal{S}$-matrix so as to restore manifest unitarity, which can be done
with an empirical algebraic procedure, exploiting the symmetry of
$\mathcal{S}$.

All this work is in progress.

\end{document}